# Perovskite Quantum Organismoids

Fan Zuo[1,*], Priyadarshini Panda[2,*], Michele Kotiuga[3], Jiarui Li[4], Min Gu Kang[4], Claudio Mazzoli[5], Hua Zhou[6], Andi Barbour[5], Stuart Wilkins[5], Badri Narayanan[7], Mathew Cherukara[7], Zhen Zhang[1], Subramanian K. R. S. Sankaranarayanan[7], Riccardo Comin[4], Karin M. Rabe[3], Kaushik Roy[2], and Shriram Ramanathan[1]

[1]School of Materials Engineering, Purdue University, West Lafayette, Indiana 47907, USA

[2]School of Electrical and Computer Engineering, Purdue University, West Lafayette, Indiana 47907, USA

[3]Department of Physics and Astronomy, Rutgers University, Piscataway, New Jersey 08854, USA

[4]Department of Physics, Massachusetts Institute of Technology, Cambridge, Massachusetts 02139, USA

[5]National Synchrotron Light Source II, Brookhaven National Laboratory, Upton, New York 11973, USA

[6]X-ray Science Division, Advanced Photon Source, Argonne National Laboratory, Argonne, Illinois 60439, USA.

[7]Center for Nanoscale Materials, Argonne National Laboratory, Argonne, Illinois 60439, USA

[*]These authors contributed equally to this work

**A central characteristic of living beings is the ability to learn from and respond to their environment leading to habit formation and decision making[1-3]. This behavior, known as habituation, is universal among forms of life with a central nervous system, and interestingly observed even in single cellular organisms that do not possess a brain[4,5]. Here, we report the discovery of habituation based plasticity utilizing a perovskite quantum system by dynamical modulation of electron localization via reversible dopant incorporation. Microscopic mechanisms and pathways that enable this organismic collective charge-lattice interaction are elucidated by a combination of first-principles theory, synchrotron investigations, *ab-initio* dynamical simulations and *in-situ* environmental breathing studies. We implement a new learning algorithm inspired from the conductance relaxation behavior of perovskites that naturally incorporates habituation and demonstrate 'learning to forget': a key feature of animal and human brains[6]. Most surprisingly, our results show that incorporating this elementary skill in learning dramatically boosts the capability of artificial cognitive systems.**

Habituation, one of the primary universal learning mechanisms, can be simply defined as the decrement in response to repeated stimuli. Habituation is seen as the simplest learning form exhibited by organisms like sea slugs[7] and fruit flies[8] to more complex living forms such as rats and humans[9,10], and is fundamental to how an organism responds and adapts to its environment thereby increasing its chances of survival. Habituation can help animals, for instance, to focus on important stimuli for novelty detection and thus can be viewed as an integral part of attention and learning[11,12] and has recently been demonstrated in the single-celled non-neural organism Phasyrum Polycephalum, commonly known as the slime mould[13]. In non-neural organisms, habituation is manifested by a change in global shape of the system (Fig. 1a). In more complex organisms that possess a nervous system, habituation has been shown to result from the decreased release of chemical transmitters at synaptic terminals[7,14]. This changes the weights of certain neural connections, a mechanism known as synaptic plasticity.

Here, we demonstrate environmental habituation-based plasticity in a perovskite oxide compound $SmNiO_3$ (SNO), which is rigorously explained by a comprehensive discussion of X-ray scattering, first-principles calculations and *ab-initio* dynamical simulations. SNO perovskite is a framework of tilted $NiO_6$ octahedra where $Sm^{3+}$ ions occupy 12-fold oxygen coordinated sites and balance the charge[15]. Hydrogen doping from the environment into lattices using catalytic electrodes occurs in reversible manner leading to massive non-linear changes in electronic properties[16,17]. Accompanied with incorporation of a proton, an electron can be injected into an unoccupied Ni $e_g$ orbital. Strong Coulomb interaction existing in $e_g$ orbitals generates a large transport gap via strong correlation effects[18,19]. As shown in Fig. 1b, after first exposure to $H_2$ (environmental stimulus 1), a significant fraction of Ni is reduced to $Ni^{2+}$, manifested by electron localization. Changing the environment (in this case by air exposure) for a short period of time followed by re-exposure to $H_2$, additional protons incorporate in SNO but with slower kinetics and keeps diminishing. While the perovskite mimics habituation, the varying conductance due to the correlated interactions shows inherent plasticity that can emulate biological synapses of neural organisms that are capable of more complex functionalities. Based on this discovery, we design a learning mechanism we term Adaptive Synaptic Plasticity (ASP) that augments traditional neural systems with a key ability of 'learning to forget' for robust and continuous learning in a dynamically evolving environment (Fig. S1).

Fig. 1c shows the underlying plasticity mechanism for memory formation and learning in the brain[20], commonly modeled with Spiking Neural Networks (SNNs)[21]. SNNs are equipped with self-learning mechanisms such as Spike Timing Dependent Plasticity (STDP) for real-time interaction with the environment[22,23]. However, in its naïve form STDP implies that any pre/post spike pair can modify the synapse, potentially erasing past memories abruptly, commonly referred to as "*catastrophic forgetting*"[24]. This phenomenon often results in severe loss of previous knowledge in a neuromorphic system that is continuously exposed to new information. Recent work suggests that the brain actively erases memories while learning to continuously process new environmental stimuli[25]. Due to limited storage space available, the brain forgets already learnt connections, gradually, to associate them with new data. Our ASP, inspired from the perovskite organismoid's variable conductance, offers a solution to the problem of catastrophic forgetting. Incorporating habituation, a *non-associative process of adaptation seen in living organisms*, into ASP learning facilitates the gradual degradation or forgetting of already learnt weights to realize new and recent information while preserving some memory about old significant data. Fig. 1d shows 'learning to forget' with ASP based weight modulation by maintaining a balance between forgetting and immediate learning to construct a *stable-plastic*[26] self-adaptive SNN for dynamic environments.

Fig. 2a experimentally demonstrates habituation in perovskite nickelate thin film devices. The initial state of the system is perturbed by exposure to a new environment (namely $H_2$). Electron doping via splitting of $H_2$ into protons and electrons results in reduction of several Ni sites to $Ni^{2+}$ that is verified by X-ray spectroscopy, causing a large decrease in conductance which can be reversed due to the weak binding of the dopant with the lattice. The temporal conductivity relaxation stems from the dynamics of surface exchange and diffusion of protons and can be modeled as an exponential relaxation that is common to thin film devices[27]. Partial reversal of doping by returning to the initial environment followed by re-exposure to $H_2$ and so forth leads to habituation manifested by a gradual reduction in response (Fig. 2a, Fig. S2). Fig. 2b shows the exponential change in conductance of the perovskite in different environments that motivates ASP learning. While the electron localization is the origin of the conductance change, the lattice breathes hydrogen as seen in the *in-situ* synchrotron studies on identical devices (Fig. 2c). To provide a microscopic understanding of doping-driven electronic structure modification, we have carried out first-principles calculations on SNO (See Fig. S4, S5 and Table S1 in Supplementary Information II for results on various magnetic orderings), as shown in Fig. 2d, primarily focusing on the addition of charge and the subsequent opening of the gap. We consider the doping of a pristine SNO state with all $Ni^{3+}$, based on observations of *Pbnm* symmetry (where all Ni sites are equivalent) at room temperature[28]. With addition of electrons one-by-one to the $\sqrt{2}$ x $\sqrt{2}$ x 2 supercell of SNO, we investigate the changes in the structure and electronic energy levels. Each added electron localizes on a $Ni^{3+}$ ion and the surrounding oxygen octahedron, shifting the lowest unoccupied orbital over 3 eV by onsite correlation to form high-spin $Ni^{2+}$ (see Table S2), which is consistent with variable-angle ellipsometry measurements[16]. This charge transfer into the Ni $e_g$ orbitals is expected from the observed changes in electron filling manifested as a shift in spectral weight in the X-ray absorption data (Fig. S6b); however, a detailed comparison requires including strong core-hole effects which is beyond the scope of density functional theory (DFT). Thus, the fully doped case (1e$^-$/Ni) shows a significantly increased band gap on the order of 3 eV. The resonant magnetic coherent soft X-ray scattering measurements (RMXS, Fig. S6c, 7) further reflect the breakdown of long-range spin order in SNO after electron doping. To then understand the migration of the protons in the lattice, we employ *ab-initio* molecular dynamics (AIMD). It is found that even at room temperature, the dopant hops from an oxygen atom to a neighboring one within $NiO_6$ octahedron (Fig. 2e). The proton is initially bound to atom O1 (Fig. 2e, i) at a distance of ~ 2.83 Å away from the O2 atom. The proton first rotates about the O1 atom (Fig. 2e, i-iii), while being bound to O1 until the O2-H distance is lowered to ~ 1.72 Å (Fig. 2e, iii). This rotation process is associated with an energy barrier of 0.27 eV, which is lower than the typical activation barriers for $H^+$ migration encountered in proton conducting perovskites (e.g., in the canonical Y-doped $BaZrO_3$, $\Delta E$ ~ 0.46e V[29]). Once the proton comes into the vicinity of O2 atom (Fig. 2e, iii), it hops over, and binds to O2 atom with a negligible energy penalty of 0.046 eV (Fig. 2e, iii-v). The proton migration between neighboring O atoms is visualized in a video in Supplementary Information V.

The conductance relaxation observed from Fig. 2b due to collective effects allows us to use the organismoid's behavior to modulate synaptic plasticity for memorization and forgetting. ASP blends non-associative habituation behavior with time-based correlation learning that helps in retention and gradual adaptation to new inputs as well as evokes competition across neurons to learn distinct patterns. We seamlessly integrate weight decay with traditional synaptic plasticity and modulate the leak rate using the temporal dynamics of pre- and post-synaptic neurons to realize habituation. While the temporal correlation helps in learning new input patterns, the retention of old data and gradual forgetting is attained with habituation. The ASP model for weight modulation with different windows for potentiation and depression based on the firing events of the post/pre neurons is shown in Fig. S8, S9 (see Supplementary Information III for details on implementation).

To demonstrate the effectiveness of the organismoid-inspired learning paradigm, against standard STDP, a fixed-size SNN (with nine excitatory neurons) was trained in a dynamic digit-recognition environment wherein digits '0' through '2' were presented sequentially with no digit re-shown to the network. Fig. 3a, b shows the representations learnt with traditional exponential STDP learning[30] against the adaptive plasticity based learning. We see that as the network is shown digit '1', ASP learnt SNN forgets the already learnt connections for '0' and learns the new input. Learning is more stable as neuronal connections corresponding to the older pattern '0' are retained while learning '1'. ASP adopts a significant- and latest-data driven forgetting mechanism (incorporated via the leak/decay phase shown in Fig. S9) wherein older digits are forgotten to learn new digits. Hence, when the last digit '2' is presented to the ASP learnt SNN, the connections to the excitatory neurons that have learnt digit '0' are forgotten in order to learn the latest digit '2' while the connections (or neurons) corresponding to recently learnt digit '1' *remain intact*, strikingly similar to human memory.

## METHODS:

**Growth of epitaxial perovskite oxide thin films.** LaAlO$_3$ (001) single crystals were used as substrate for epitaxial growth of SNO thin film by magnetron co-sputtering of metallic nickel in direct-current mode (at power of 75 W) and samarium targets under radio-frequency mode (at power of 150 W) in 5 mTorr of argon and oxygen mixture flowed at rate of 40/10 standard cubic centimeters per minute (sccm). The as-sputtered sample was then transferred into a custom-built high pressure vessel and annealed under 100 bar of pure oxygen at 500 °C for 24 hours in a tube furnace.

**Electrical characterization.** *In situ* temporal resistance measurements were done in a sealed custom-designed chamber equipped with a gas flow controlling system, from which we could switch the chamber inner atmosphere between various environments. In this experiment, we use 5% H$_2$ balanced by 95% argon as the stimulus environment, and air as the recovery environment. Four 100 nm Pt strips (0.5 mm X 5 mm) were deposited onto the top of epitaxial SNO thin films by E-beam evaporation. The strips function as catalyst to split H$_2$ into protons and electrons for the electron doping of SNO, and electrical contacts for the resistance measurements as well. The distance between each Pt strip is 1 mm. The experiments were conducted at 50 °C to maintain a steady temperature throughout the studies. Resistance was calculated from current-voltage curves swept between -0.1 V to 0.1 V using Keithley 2635A instrument. The initial chamber atmosphere was air and Pt/SNO sample was placed onto the sample stage for 30 min to reach stable temperature. Real time resistance testing, was conducted using a custom LabVIEW code. To avoid temperature fluctuation during gas switching, the 5% H$_2$/95% Ar was flowed to the chamber at a moderate flow rate of 30 sccm. After 30 min reaction, the gas was switched back to air by flowing dry air at the same flow rate (30 sccm). 10 min later, gas was changed to 5% H$_2$/95% Ar again, and this H$_2$-Air cycle was repeated. The relaxation of conductivity after exposure to environment is fitted with the formalism to study conductivity relaxation in thin film devices[31,32], $(\sigma_0 - \sigma_t)/\sigma_0 = C - A_1 \exp(-k_1 t) - A_2 \exp(-k_2 t)$, where $\sigma_0$, and $\sigma_t$ represent initial and dynamical conductivity, respectively, $k_1$ and $k_2$ depicting the relaxation kinetics.

***In situ* synchrotron X-ray diffraction.** Synchrotron X-ray diffraction measurements of the SNO devices were conducted at an insertion device beamline, 12ID-D at the Advanced Photon Source, Argonne National Laboratory, on a six-circle Huber goniometer with an X-ray energy of 20 keV using a pixel array area detector (Dectris Pilatus 100 K). The X-ray beam had a flux of $10^{12}$ photons per second. The $q_z$-scan (L-scan) was obtained by removing the background scattering contributions using the two-dimensional images. For *in situ* X-ray diffraction testing, Pt/SNO sample was placed in a testing compartment sealed by Kapton tape. The testing condition followed the atmosphere progression mode shown in Fig. 2a.

**Resonant magnetic coherent soft X-ray scattering (RMXS).** Resonant magnetic coherent soft X-ray scattering study was performed at the beamline 23-ID-1 of the National Synchrotron Light Source II (NSLS-II), at Brookhaven National Laboratory. All data were collected using horizontally polarized light and a vertical scattering geometry, with photon energy tuned near the Ni-$L_3$ absorption edge. The probing geometry is illustrated in Fig. S6a. The pristine SNO thin film is patterned with Pt stripes and hydrogen is intercalated to yield electron-doped regions (H-SNO) of width 0.1 mm. The magnetic scattering signal is measured by a two-dimensional CCD positioned 33 cm from the sample, while the X-ray absorption is collected in total fluorescence yield also using the CCD (away from structural or magnetic reflections). In order to reach the magnetic reflection, at Q = (1/4,1/4,1/4) the sample was oriented so that the scattering plane is spanned by crystal vector [110] and [002]. The film was illuminated by a coherent X-ray beam whose coherent fraction is selected by a 10 μm diameter pinhole in close proximity to the sample. The measurements were performed at ~20 K which is well below the Neel temperature of SmNiO$_3$ (~ 200K).

**First-principles calculations for SNO electronic structure**. First-principle calculations were carried out within the density functional theory (DFT)+U approximation with the Vienna Ab-initio Simulation Package (VASP) code[33,34] using the projector augmented plane-wave PAW method of DFT[35] and the supplied pseudopotentials: Sm_3 (valence: *5s$^2$5p$^2$6s$^2$4f$^1$*), Ni_pv (valence: *3p$^6$4s$^2$3d$^8$*) and O (valence *2s$^2$2p$^4$*). To treat the exchange and correlation, the PBE functional was used within the generalized gradient approximation (GGA)[36,37] and the rotationally invariant form of DFT+U of Liechtenstein et al.[38] with U = 4.6 eV and J = 0.6 eV. For structural determination of pristine SNO, we started with the Materials Project structure[39] added a small monoclinic distortion (ß ≈ 90.75°) and allowed the cell and ionic positions to relax until the forces were less than 0.005 eV/Å on each ion. All calculations were carried out with the tetrahedral method with Blöchl corrections[40], a 6x6x4 Monkhorst-Pack *k*-point mesh for the $\sqrt{2}$ x $\sqrt{2}$ x 2 supercell, and a plane-wave energy cutoff of 500eV. When simulating electron-doping, extra electrons were added to the calculation with a positive background compensation charge. For SNO, we added 1, 2, 3 or 4 electrons to the monoclinic $\sqrt{2}$ x $\sqrt{2}$ x 2 supercell with a G-type magnetic ordering, resulting in an electron doping concentration of 1/4, 1/2, 3/4, 1 $e^-$/Ni, respectively. In each case, we allowed the internal ionic positions to relax, using the same force tolerance as before, while keeping the volume and cell shape unchanged. When studying effect of magnetic order on the fully doped case (Fig. S4), both the ionic positions and the lattice parameters were relaxed.

***Ab-initio* molecular dynamics simulations.** We performed *ab-initio* molecular dynamics (AIMD) simulations within GGA with Hubbard correction using the PAW formalism as implemented in VASP [33,34]. The computational supercell consists of 4 unit cells of monoclinic SNO (2x2x1 repetitions of unit cell; 80 atoms). Periodic boundary conditions are employed along all directions. The exchange correlation is described by the PBE functional[36,37], with the same pseudopotentials as used in the electronic structure calculations. The plane wave energy cut-off is set at 520 eV. The Brillouin zone is sampled at the Γ-point only. Using AIMD simulations in the isobaric-isothermal (NPT) ensemble, we first thermalize the SNO computational supercell at various temperatures ranging from 300 K-800 K and zero external pressure for 10 ps using a timestep of 0.5 fs. During these simulations, the cell volume, cell shape, as well as the atomic positions are allowed to vary via the Parrinello-Rahman scheme[38]; the temperature conditions are maintained by using a Langevin thermostat[41]. Next, we insert a proton within the thermalized SNO (at a given temperature), such that it is at a distance of 0.98 Å away from an arbitrarily chosen O atom. Note that we ensure supercell neutrality upon addition of the proton via a background negative charge. To monitor the diffusion of the inserted proton (Movie 1, Supplementary Information V), we perform AIMD simulations at constant volume (and shape) and temperature (i.e., NVT ensemble) for an additional 10 ps. For these AIMD simulations of doped SNO, constant temperature conditions are maintained via Nose Hoover thermostat[41] as implemented in VASP.

**Simulation methodology for spiking neural networks**. The ASP learning algorithm was implemented in BRIAN[42] that is an open source large-scale SNN simulator with parameterized functional models (Leaky-Integrate-and-Fire) for spiking neurons. We used the hierarchical SNN framework (Fig. S8) to perform digit recognition with the MNIST dataset[43]. The network topology and the associated synaptic connectivity configuration were programmed in the simulator. The spiking activity (or time instants of spikes) of pre- and post-neurons were monitored to track the corresponding pre/post synaptic traces that were used to estimate the weight updates in the recovery/decay learning phase of ASP.

**Supplementary Information** is available in the online version of the paper.

**Acknowledgements** Financial support was provided by ARO, AFOSR, C-SPIN, one of the six centers of STARnet, a Semiconductor Research Corporation Program, sponsored by MARCO and DARPA, by the Semiconductor Research Corporation, the National Science Foundation, Intel Corporation and by the Vannevar Bush Faculty Fellowship. We acknowledge the Center of Nanoscale Materials User Facility at Argonne National Laboratory. M.K. and K.M.R. acknowledge support from Office of Naval Research grant N00014-12-1-1040. This research used resources of the Advanced Photon Source, a U.S. Department of Energy (DOE) Office of Science User Facility operated for the DOE Office of Science by Argonne National Laboratory under Contract No. DE-AC02-06CH11357. This research also used resources at the 23-ID-1 (CSX-1) beam line of the National Synchrotron Light Source II, a U.S. Department of Energy (DOE) Office of Science User Facility operated for the DOE Office of Science by Brookhaven National Laboratory under Contract No. DE-SC0012704. M.G.K. acknowledges a Samsung Scholarship from the Samsung Foundation of Culture. We acknowledge C.Huang for the assistance with electrical measurements.

**Author Contributions** F.Z., P.P., K.R. and S.R. conceived the study. F.Z. fabricated the SNO device and performed the initial experiments. P.P. designed and investigated the learning models. F.Z. and H.Z. performed the X-ray diffraction characterizations. M.K. and K.M.R. conducted the first-principles calculations, B.N., M.C. and S.K.R.S.S performed the *ab* initio simulations. J.L., M.G.K., A.B., C.M., S.W. and R.C. performed the resonant magnetic coherent soft X-ray scattering. Z.Z. grew the SNO thin film. F.Z., P.P., M.K. and S.R. wrote the manuscript. All authors discussed the results and commented on the manuscript.

**Author Information** Reprints and permissions information is available at www.nature.com/reprints. The authors declare no competing financial interests. Readers are welcome to comment on the online version of the paper. Correspondence and requests for materials should be addressed to S.R. (shriram@purdue.edu) or K.R. (kaushik@purdue.edu).

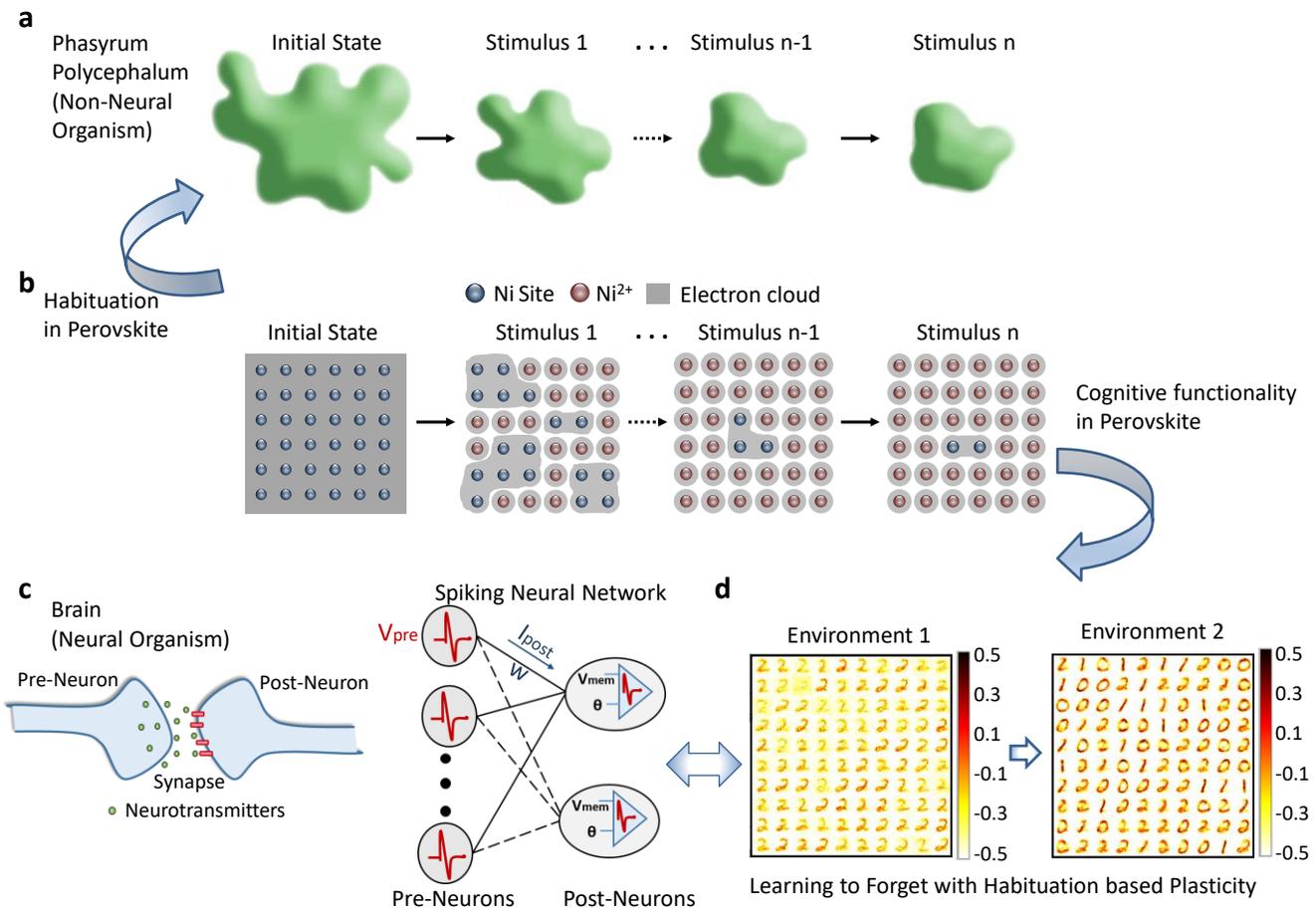

**Figure 1 | Quantum Material showing habituation behavior observed in neural and non-neural organisms. a**, Non-associative habituation learning observed in Physarum polycephalum. When exposed to stimulus, a diminished response is observed indicative of habituation. **b,** Schematic showing the habituation process in a perovskite (SNO). Between repeated stimuli ($H_2$), the dynamics of carrier localization subsides, showing *both* non-neural habituation and neural synaptic plasticity. **c,** Associative spike timing based learning observed in a biological neural system (brain) responsible for memory formation. In the brain, synaptic plasticity is modulated by chemical transmitters, and is a function of the relative timing difference between the post and pre-neuronal spikes. The biological neural system is implemented as a Spiking Neural Network (SNN) that consists of a fully-connected array of pre-neurons and post-neurons. The pre-neuronal voltage spike ($V_{pre}$) is modulated by the synaptic weight (w) to generate the resulting post-synaptic current ($I_{post}$). The post-neuron integrates the current that results in an increase in its membrane potential ($V_{mem}$) and spikes when the potential exceeds a certain threshold (θ). **d,** In environment 1, the SNN was presented with different images of digit '2' and learnt several patterns corresponding to the given image. In environment 2, the SNN was presented with images of digits '0' and '1'. Incorporating habituation based non-associative learning with standard associative STDP enables the SNN to learn new patterns without catastrophic forgetting in a resource-constrained dynamic input environment.

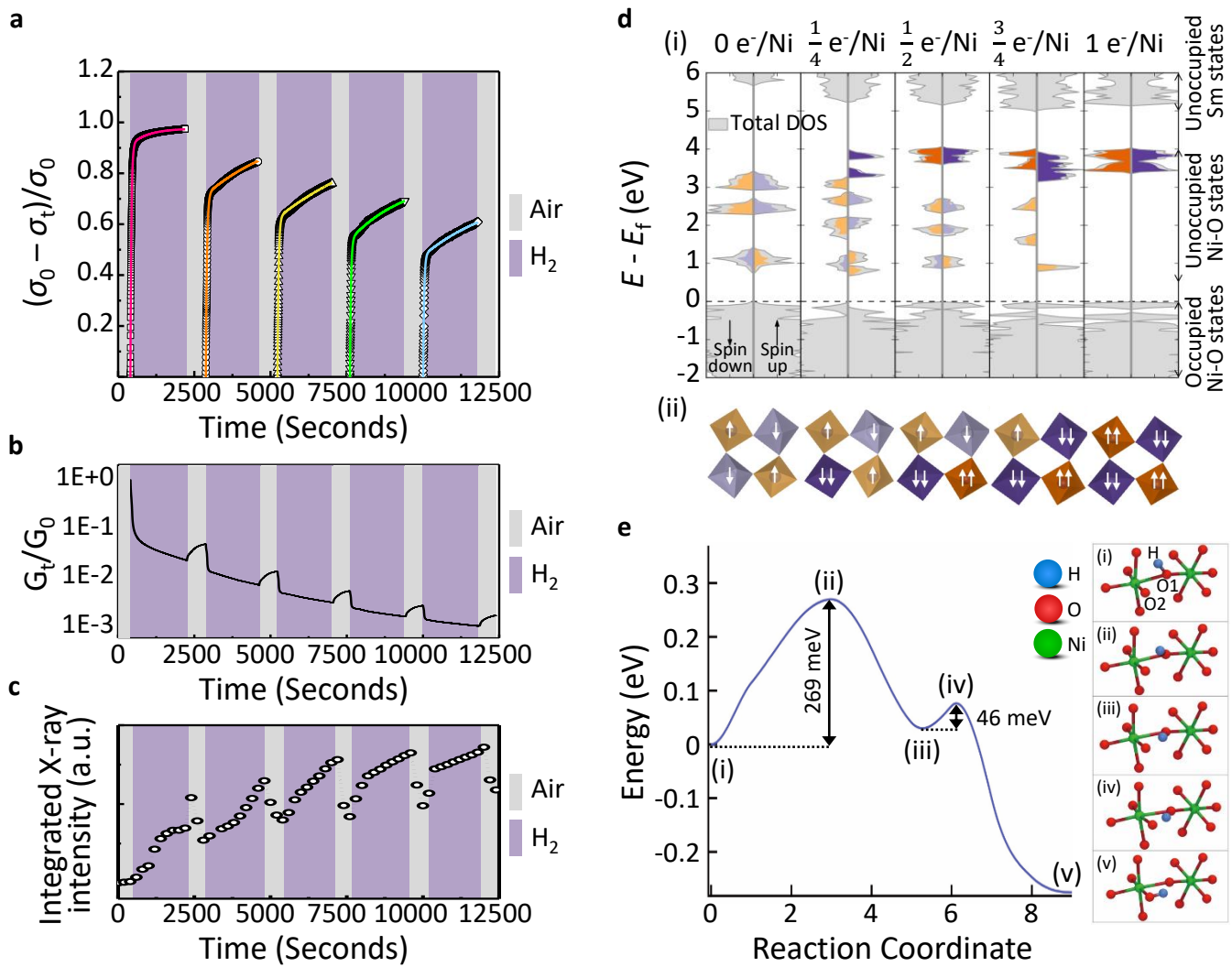

**Figure 2 | Mechanism of habituation in a perovskite nickelate. a,** *In situ* visualization of habituation phenomenon, i.e. exponential decrease of conductivity change upon environmental exposure (the dots represent the experimental data and the solid lines are fits.). $\sigma_0$ and $\sigma_t$ are initial and dynamical conductivity, respectively. **b,** The conductance changes in response to different environments (decrease in $H_2$ and increase in Air) showing inherent plasticity similar to what is observed in biological synapses. $G_0$ and $G_t$ represent initial and dynamical conductance, respectively. **c,** Structural lattice breathing monitored by *in-situ* synchrotron X-ray diffraction. The integrated intensities of x-ray diffraction peak at $q_z = 2.98$ Å$^{-1}$ related to H-SmNiO$_3$ (H-SNO) are shown (see Fig. S3). **d,** First-principles calculation of electron doped SNO. (i) shows density of states (DOS), in gray, at different doping levels from 0-1 added e$^-$ per Ni site. The unoccupied projected DOS (PDOS) on each nickel site is shown in orange and purple. The difference in the total DOS and the PDOS is due to the strong hybridization of the Ni and O states resulting from the covalent nature of the NiO$_6$ octahedra. (ii) shows the occupied Ni $e_g$ levels for the corresponding doping levels. Same color legend is used and the darker colors indicate Ni with two occupied $e_g$ states. **e,** Atomic-scale pathway, and the associated energy barriers for proton migration between two neighboring O atoms labeled as O1 and O2 in (i) within a NiO$_6$ octahedron in a monoclinic SNO crystal. The potential energy along the most preferred diffusion pathway (as obtained from nudged-elastic band DFT calculation) is shown on the left, while selected configurations along this pathway labeled (i)-(v) are depicted on the right.

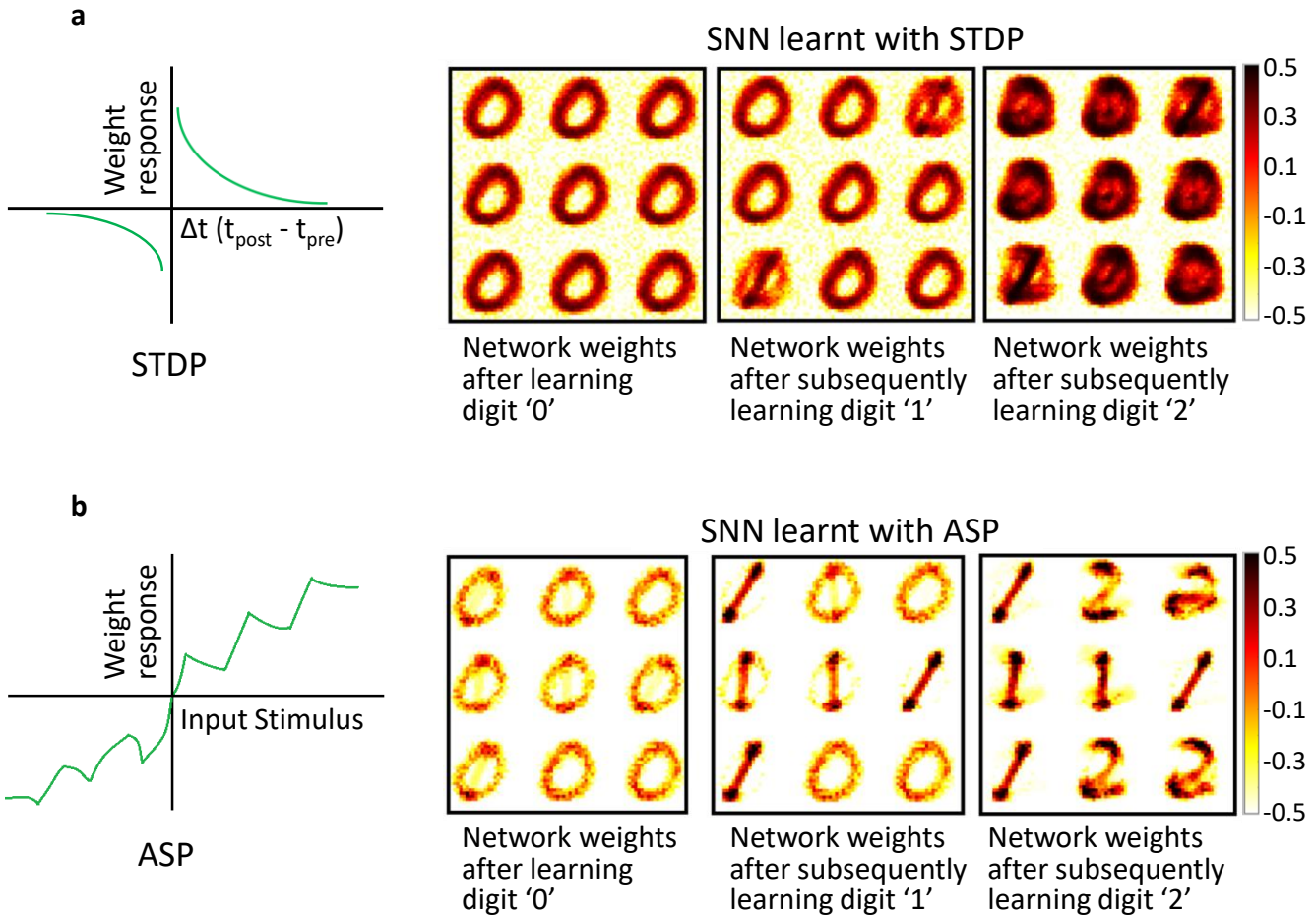

**Figure 3 | Learning by forgetting. a, b,** Digit representations learnt with digits '0' through '2' shown sequentially to an SNN (with 9 excitatory neurons) trained with standard STDP (**a**) and ASP that integrates habituation (**b**). Presenting the digits one-by-one sequentially i.e. first all the images for digit `0' followed digit `1', and so on can be treated as a dynamic learning environment. No particular digit instance or class is re-shown to the network. SNN trained with STDP tried to learn the new digit representation (for instance, digit '1') while retaining a portion of the old data (for instance, digit '0'). However, fixed network size and absence of data reinforcement (i.e. no any old data or digit showing with the new data) resulted in accumulation causing new weight updates to coalesce with already learnt patterns rendering the network incapable of categorizing the digits. In sharp contrast, ASP learnt SNN, with identical resource constraints in place, gracefully forgets old patterns and adapts to learn new inputs effectively without catastrophically erasing old data. Fig. S9 shows the representations learnt for a larger network when all digits '0' through '9' are presented.

**Supplementary Information:**

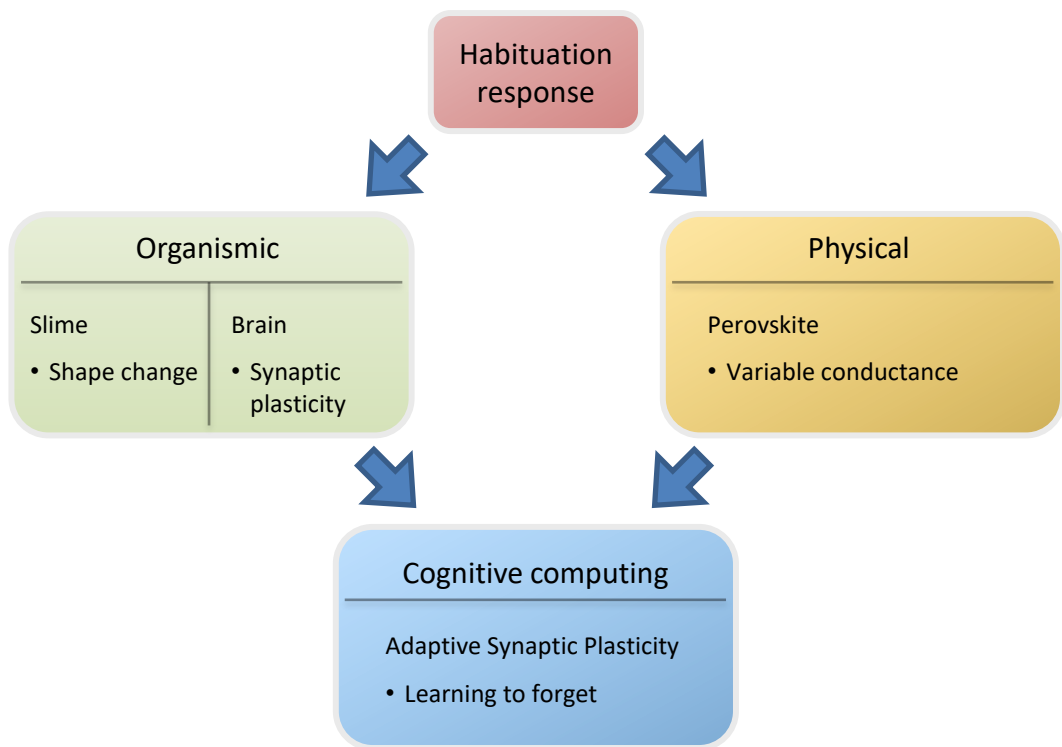

**Figure S1 | Habituation response in the perovskite and learning mechanism in neural/non-neural organisms.** Habituation is a ubiquitous behavior present across the phyla of living beings that help organisms to learn and adapt to different aspects of the environment. It has been demonstrated to cause short-term and long-term potentiation of synaptic connections (or synaptic plasticity) that is key to memory formation in neural organisms. In non-neural organisms such as slime, habituation is seen as a change in its shape in response to different environments. The perovskite's non-linear response to the environment ($H_2$ and Air in this study) with varying conductance mimics simple adaptation behavior and motivates the Adaptive Synaptic Plasticity (ASP) learning (see Supplementary Information I and Fig. S9). In ASP, we incorporate habituation by weight leaking coupled with traditional spike timing correlation to demonstrate 'learning to forget' for robust and stable learning of artificial neural systems.

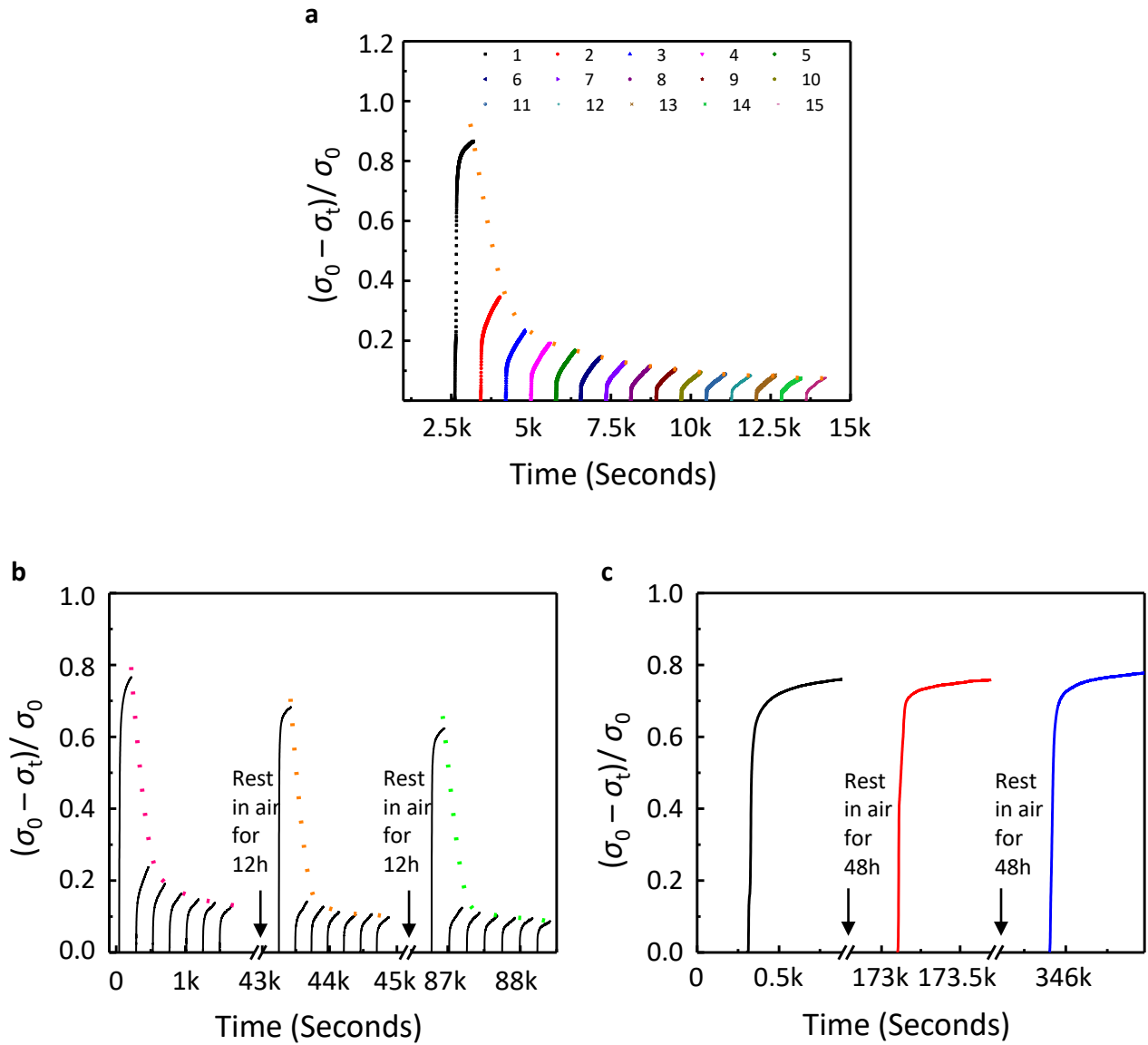

**Figure S2. Habituation-forgetting-habituation and complete recovery behavior of SNO device.** In organisms, if the stimulus is withheld for extended period of time, the original state can be recovered. Subsequent exposures to the environment will result again in habituation. Strikingly identical to this behavior, our nickelate devices can be made to forget previous exposures to hydrogen by resting in air. **a**, A set of experiments on a device after it was recovered by an air anneal then followed by testing for 15 cycles. The continuous diminished response demonstrates habituation behavior. **b**, After seven cycles of $H_2$/Air treatments, the SNO was left in air for 12h. The device started to recover and approached its original state. With the same manner of $H_2$/Air exposure again, the habituation behavior could be reproduced. This forgetting-habit forming process was repeated by resting the SNO device in air for another 12h and re-exposure to $H_2$/Air. **c**, non-habituation by fully recovering SNO. It is worth noting that if the $H_2$ treatment was followed by an extended exposure of the nickelate device in air for 48h, no habituation phenomenon would be present, again similar to what is observed in experiments conducted on organisms. The dotted conducting line in (**a**) and (**b**) indicated the trend of diminished response. These experiments were conducted in a manner identical to numerous experiments conducted on organisms in the biology literature and are all cited in the manuscript and Supplemental Information files.

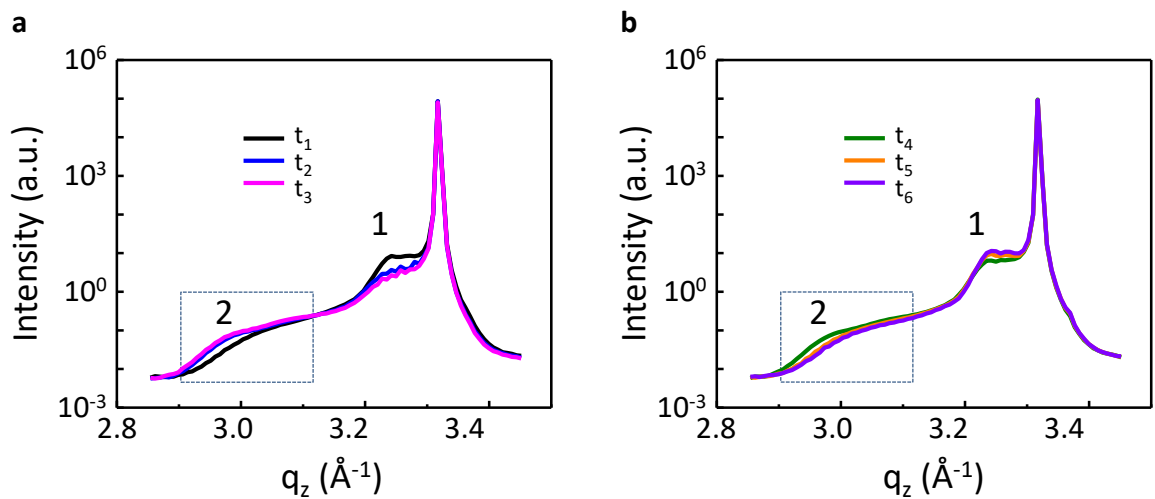

**Figure S3 | A set of representative *in situ* synchrotron X-ray diffraction patterns monitoring structural evolution upon breathing in H₂ and air sequentially. a**, X-ray diffraction patterns in H$_2$ environment. SNO thin film and substrate LAO are indexed in pseudocubic lattice system. SNO (002) peak (labeled 1) appears close to LAO (002) peak. The peak at q$_z$ = 2.98 Å$^{-1}$ (peak 2) is related to H-SNO[1]. Peak 2 is broad and weak due to low H-SNO phase content and interstitial disorder due to dopant incorporation. It is clearly seen that when treated with H$_2$, peak 1 drops with exposure time (t$_1$→t$_2$→t$_3$), while peak 2 increases with longer exposure, indicating H-SNO phase emergence. **b**, X-ray diffraction patterns in air environment. The gas environment was switched to air and structural evolution was monitored. It is evident that no new peaks are present. An opposite trends is observed, i.e. the SNO phase is restored and H-SNO phase diminished when breathing in air. The areas of peak 2 for all measured data during H$_2$/air breathing are calculated via integration of the region shown in dotted line box and plotted in Fig. 2c.

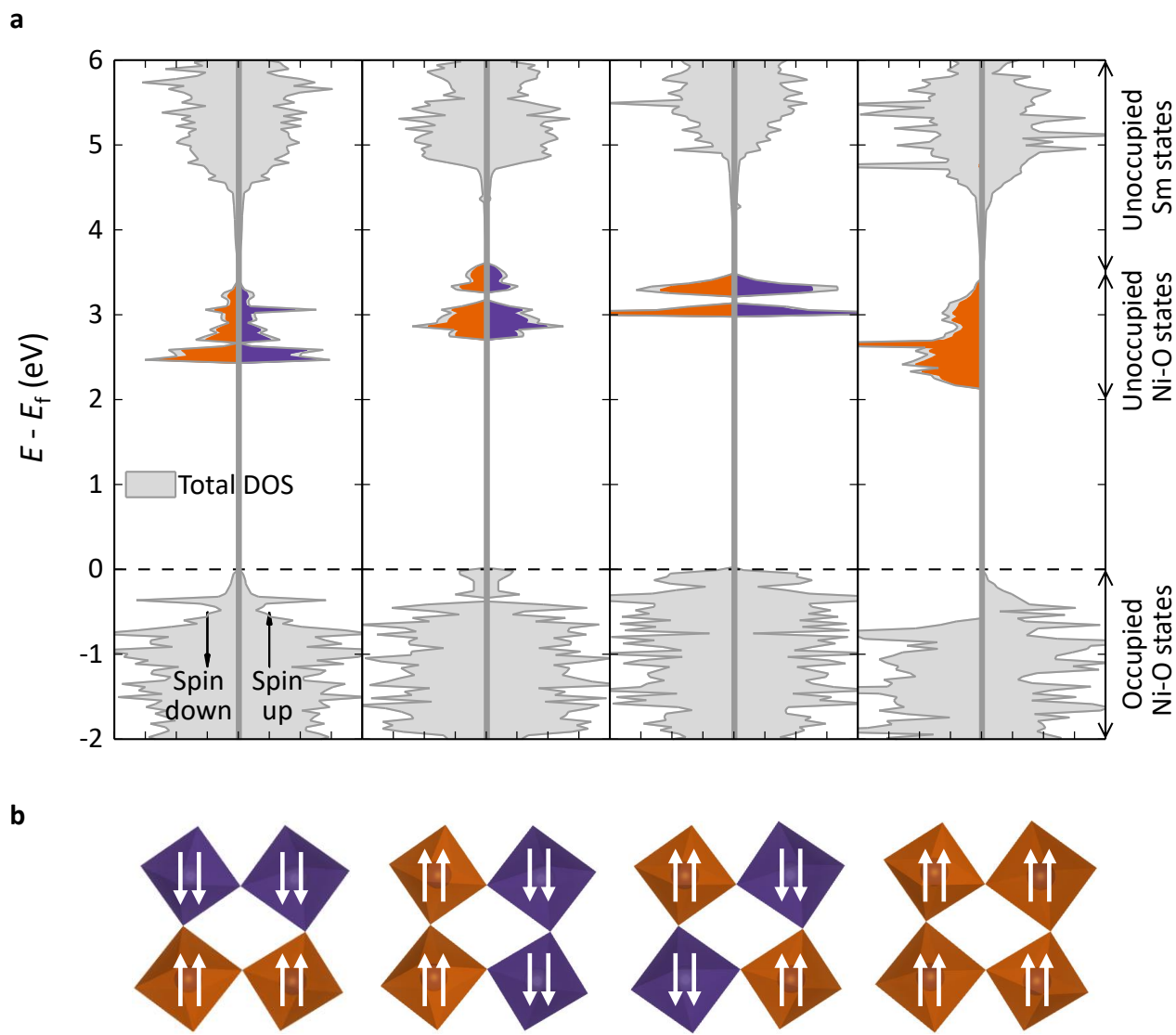

**Figure S4 | Density of States (DOS) based on different magnetic orderings of doped SNO. a**, DOS for fully doped (1e⁻/Ni) SNO - all $Ni^{2+}$ - with various magnetic orderings. Despite the magnetic ordering, all band gaps are on the same order. The structures are fully relaxed and the $Ni^{2+}$ are in high spin configurations, which is favored over the low spin configuration due to the Hund's coupling. Colored states are the unoccupied Ni projected DOS (PDOS), corresponding to $e_g$ states. **b**, The occupied Ni $e_g$ electrons on each Ni site for each magnetic ordering. The colors of the octahedra correspond to the colored PDOS in (**a**).

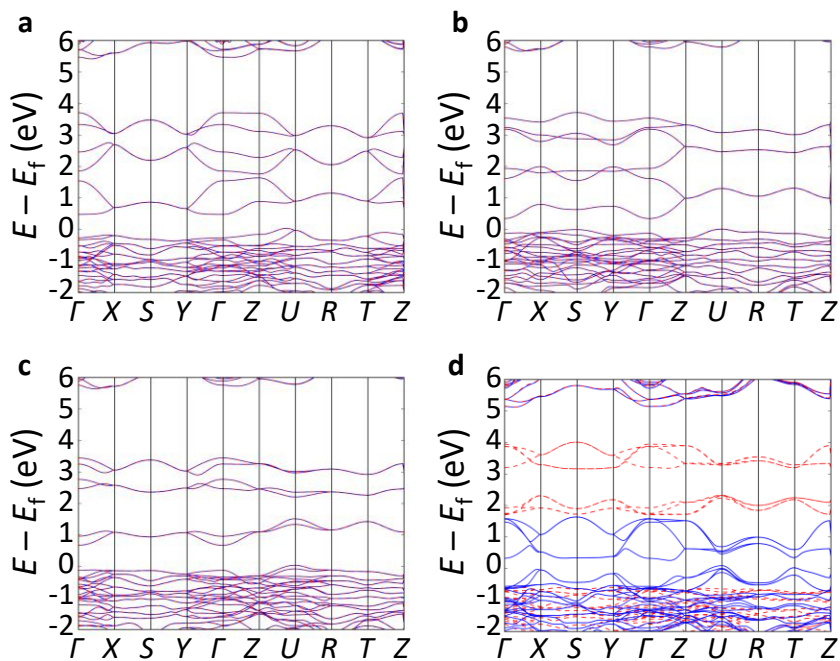

**Figure S5 | Band structure of the monoclinic structure with Jahn-Teller distortion. a-d** show the band structure with (**a**) antiferromagnetic orderings with layers perpendicular to the z-axis (A-type); (**b**) with columns along the z-axis, forming a checkerboard in the xy-plane (C-type); (**c**) with a three dimensional checkerboard (G-type); and (**d**) ferromagnetic ordering (F-type). For A-, C-, and G-type, the bands are spin degenerate. The two highest valence bands are the 4 occupied Ni $e_g$ electrons (one on each Ni site). The 6 bands between 0 and 4 eV are the unoccupied Ni $e_g$ states. For F-type, the spin up bands are shown in solid blue, and the spin down bands are shown in dashed red. The four highest valence bands are the 4 occupied Ni $e_g$ electrons (one on each Ni site). All of these bands are spin up (the four bands can be clearly seen at the "U" point), leaving 8 unoccupied spin down bands and 4 spin up. The bands between 0 and 4 eV are the unoccupied Ni $e_g$ states.

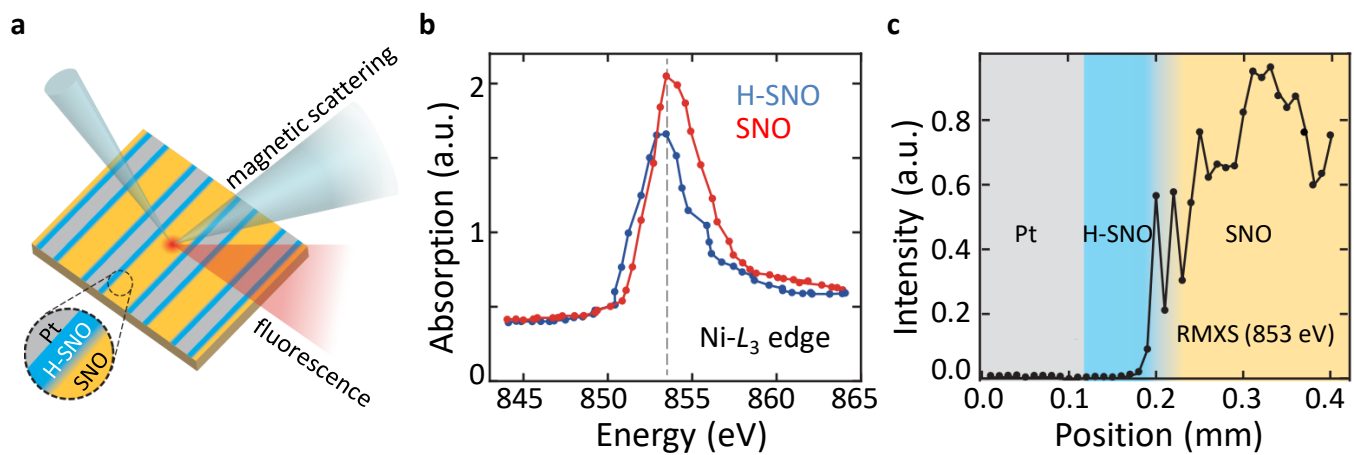

**Figure S6 | Resonant magnetic X-ray scattering (RMXS) study of SNO after breathing in $H_2$. a**, The experimental setup at beamline 23-ID-1 (CSX-1) at the NSLS-II. In the SNO region (orange), the sample is identical to the pristine (un-doped) SNO. The film regions close to Pt bars (gray) are hydrogenated and marked as 'H-SNO' (blue). **b**, Ni-$L_3$ X-ray absorption profiles of H-SNO (blue) and SNO (red). The absorption peak at 853 eV in SNO corresponds to the Ni-$L_3$ edge as noted in literature[2]. When doped with electrons, the absorption edge shifts to lower energy, and develops a shoulder on the high energy side. This shift in spectral weight is indicative of an increase of electron filling on the Ni site suggesting doping-induced changes in the orbital filling and electronic structure. **c**, Spatial RMXS study of SNO. Position-dependent magnetic peak intensity was collected at 853 eV (dashed line in **b**). The magnetic reflection is only present in the SNO region.

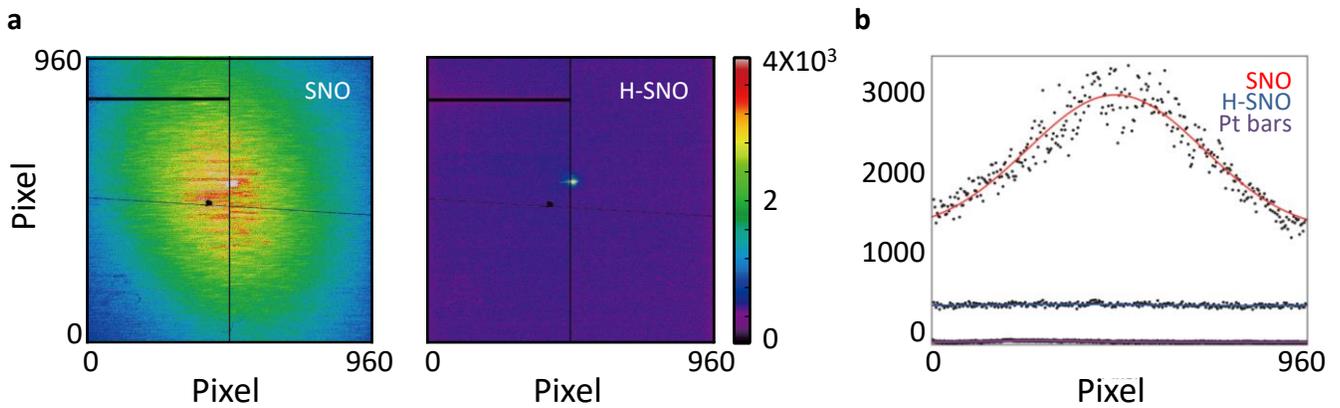

**Figure S7 | Typical magnetic scattering signal for SNO and H-SNO region of sample after breathing in $H_2$. a**, Detector images of the magnetic reflection peak from different regions of the sample. A diffuse magnetic peak centered at the wavevector **Q** = (1/4,1/4,1/4) can be clearly seen when the undoped SNO is illuminated (left), while no peak was found in H-SNO (right). Both measurements have been performed at 20 K, well below the Neel temperature (200 K) of $SmNiO_3$. The black line crossing the center of CCD is the shadow of wires holding beam stop (shown as a black spot near center). **b**, Magnetic reflection signal for three different regions of the sample (SNO, H-SNO, and Pt bar), with Gaussian fit (plus baseline) overlaid. To obtain the RMXS intensity, we integrated the CCD signal along vertical slices (with 50 pixels lateral width) through the peak center in order to average out the intensity fluctuations (speckle pattern) arising from domain interference (**b**). When the beam is on the Pt bar, no signal can be measured, due to the opacity of the heavy metal layer to soft X-rays. The fluorescence background increases in the H-SNO region, where, however, no magnetic scattering can be detected. The magnetic reflection only appears in the undoped SNO region. All linecuts are fitted using a Gaussian lineshape with a uniform background. The position-dependent magnetic peak intensity can then be extracted as shown in Fig. S6c.

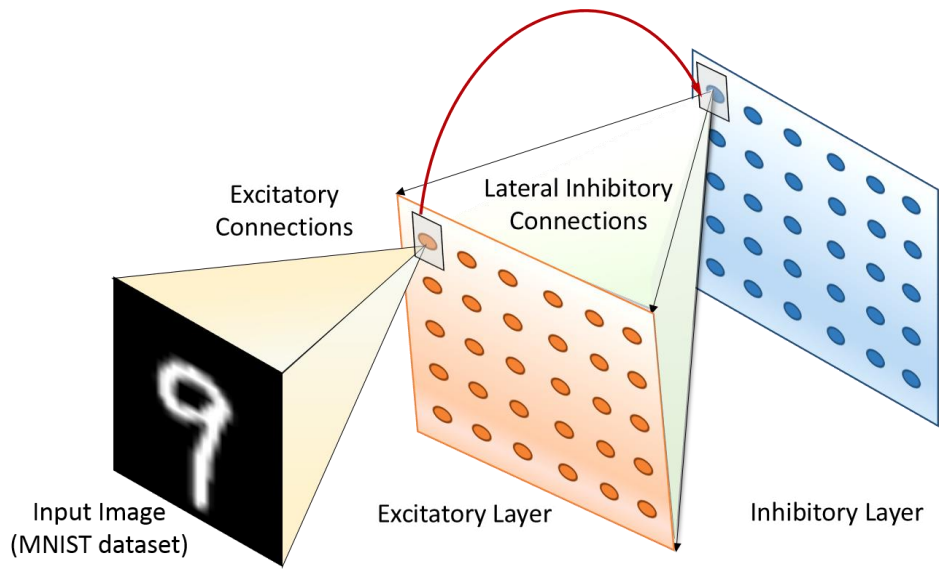

**Figure S8 | SNN topology for pattern recognition consisting of input, excitatory and inhibitory layers arranged in a hierarchical fashion.** The topology consists of an input layer followed by excitatory and inhibitory layers. The input layer contains 28x28 pixel image data (with one neuron per image pixel) from MNIST dataset[3]. Each input pattern or image is converted to a Poisson spike train based on the pixel intensities of the images in the dataset. The input layer is fully connected to the excitatory neurons, that are connected to the corresponding inhibitory neurons in a one-to-one manner. Each of these neurons inhibit the excitatory layer neurons except the one from which it receives the forward connection. This connectivity structure provides lateral inhibition that limits the simultaneous firing of various excitatory neurons in an unsupervised learning environment, promotes competitive learning causing them to learn different input patterns from each other. Besides lateral inhibition, we employ an adaptive membrane threshold mechanism called homeostasis[4] that regulates the firing threshold to prevent a neuron to be hyperactive. It equalizes the firing rate of all neurons preventing single neurons from dominating the response. During learning, the excitatory synaptic weights from the input layer to each excitatory neuron are modulated to learn a particular input digit using the learning rule. Towards the end of the learning phase, the weights (or excitatory connections) that are randomly initialized eventually learn to encode a generic representation of the digit patterns. Specifically, the weights fanning out of the higher-intesity (or white) pixel regions will get potentiated while the weights from the low-intensity regions on the input image will be depressed during the learning phase. Correspondingly, the color-map figures shown in Fig. 1d, Fig. 3 and Fig. S9b represent the weight values learnt corresponding to each excitatory neuron when learning stops.

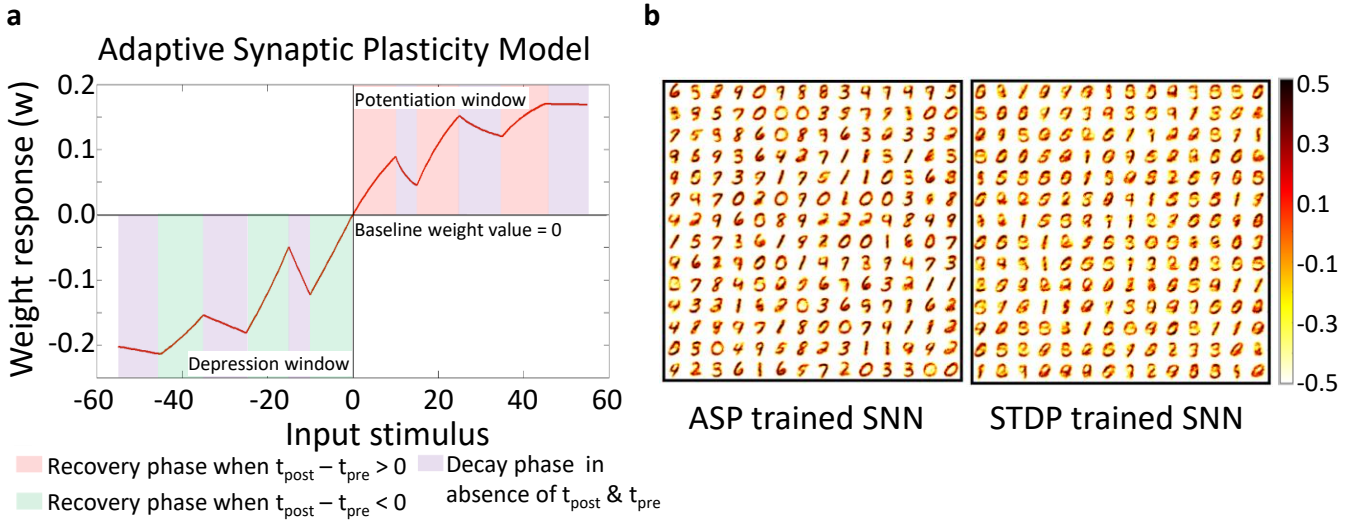

**Figure S9 | Adaptive Synaptic Plasticity learning for weight modulation in a SNN. a,** When a spiking activity is observed at the post/pre neuronal synaptic terminal, the recovery phase begins. This phase involves an exponential increase (potentiation) or decrease (depression) based on the temporal difference between the spiking activities of the pre- and post- neurons. The decay or forgetting phase in a synaptic weight ensues when there is no spiking activity (or input stimulus) observed at both post/pre neuron. The weights are dynamic during the training phase when input patterns are presented. The leak dynamics determine which post-neuronal connections (that have learnt old or insignificant data) should be forgotten to learn the new data. Whilst, the recovery phase potentiation/depression is geared towards making synaptic weight updates to learn a generic representation rather than learning specific training patterns. For instance, the weights of an excitatory layer post-neuron learning a digit '2' should spike for different instances of '2' so that it learns a more generic representation rather than just mimicking a specific instance. Thus, synaptic depression (based on *spike timing correlation*) and leak (based on *habituation*) have different roles in ASP learning. ASP incorporates the significance of the inputs to modulate the weights (see Supplementary Information III for details on implementation). That is, the weight updates during recovery phase are more prominent for frequently spiking input neuron. Also, the leak rate during the decay phase is varied taking into account the post-synaptic or excitatory neuron's spiking activity and membrane threshold $v_{thresh} + \theta$, such that recent input patterns do not overwrite old but significant data. The leak rate decreases as the weights become more and more prominent (either in potentiation or depression window), that is basically habituation. This behavior helps to retain significant information (corresponds to weights with higher negative/positive values) while forgetting (or leaking) the weights corresponding to insignificant information. **b,** To show the effectiveness of the proposed learning model for larger problems, we trained an SNN of 200 excitatory neurons with STDP and ASP in a dynamic environment when all digits '0' through '9' are presented sequentially. To ensure that the earlier digits are not completely forgotten, the number of training instances of each digit category were arranged in a decreasing order i.e. digit '0' had more training instances than digit '1' and so on. So, the network will try to retain more significant data while learning recent patterns. It is clearly seen that the SNN learnt with our proposed ASP encodes a better representation of the input patterns in comparison to the standard STDP trained network. In fact, the network is able to represent all digits. Without habituation in STDP trained SNN, most of the representations are illegible due to substantial overlap. As noted in Supplementary Information IV, ASP learning can also be naturally integrated with filamentary switch or spin-based devices.

## I. Additional notes on variable conductance and dynamics

Habituation is referred to as '*diminishing of response to a frequently repeated stimulus*' in the organismic biology literature[5]. We explain the principles further in this section to enable a general platform for materials design and illustrate the experimental envelope and dynamics. To achieve plasticity in the perovskite, we vary the environment dynamically without allowing the system to achieve equilibrium. For instance, if we simply expose the perovskite at 50°C to hydrogen indefinitely, the resistance will increase to a self-limiting value and saturate over a period of several hours. Instead, we perturb the system well before saturation and then re-expose to the environment before complete reversal to the original state. In this manner, the environment is dynamically modulated over appropriate time scales quite similar to studies conducted on organisms. The perovskite retains memory of the previous exposure since not all of the dopant has left the lattice and is therefore continuously modified, leading to diminished response. The perovskite nickelates show large non-linear changes in conductance upon electron doping due to strong correlations in the partially filled $e_g$ Ni orbitals, enabling the mimicry of environmental plasticity. Other materials that show non-linear changes of functional properties in response to external stimuli may be considered to further investigate similar organismic behavior. Similarly, for reversibility, weak binding of the dopant to the lattice and escape back into the environment is important. The time scale in our experiments was chosen in part to be in the range of experimental studies in biology, and is also close to what is noted for decision making in ant colonies, bees and related species where environmental chemical traces (i.e. diffusion of gases or scents over sensory distances[6,7]) are used for foraging food or collective locomotion to illustrate proof-of-principle. As a comparison, the response of neural connections in the brain is of the order of milliseconds and electronic memory is of the order of sub-microseconds. To mimic such faster time scales, instead of varying the gas-phase species in an environmental chamber (like we have done at a synchrotron beamline), one can use thin film solid or liquid electrolytes interfaced with the perovskite or other materials systems and rapidly move protons, oxygen, lithium or other ions into and out of the habituating material. Since ions are charged and electrons can be reversibly anchored to the partially filled d-orbitals, electric fields can be used to operate these devices that can be integrated onto circuits[8-10]. The varying conductance of the perovskite organismoid indicates an inherent plasticity that can be used for creating artificial cognitive systems. This motivated us to design ASP that incorporates habituation with traditional spike-timing correlation based learning. It is implemented by modulating the exponential leak rate of the weights based on the significance of the incoming inputs and is critical for 'Learning to forget' in a dynamic environment. As explained in a later section, this model is also compatible with other classes of oxide-ionic devices that incorporate filamentary switching or spin-based devices.

## II. First-Principles calculation of SNO band structure:

The undoped (pristine) calculation are carried out on monoclinic SNO with a Jahn-Teller distortion, relaxed from a *Pbnm* structure and freezing in a monoclinic distortion with β ≄ 90. This structure (space group $P2_1/n$) is a √2 x √2 x 2 supercell with two inequivalent Ni sites. Considering 10 atom/cell magnetic orderings for the monoclinic structure, we found the energies of all these relaxed structures to be within 0.1 eV/Ni of each other with band gaps all smaller than 1 eV (See Table S1 for further details). Consequently, the specific choice of the underlying magnetic structure in the spin-polarized DFT+U is not expected to influence appreciably the evolution of the band structure with doping. Previous computational work on rare earth nickelates, employing both dynamical mean field theory (DMFT) and DFT+U, has shown that the insulating phase exhibits a disproportionated structure, which is often exaggerated by DFT+U[11]; furthermore, room temperature experiments show only a small[12] or no[13] disproportionation. For each added electron, its localization can be observed through the magnetic moment of the Ni, the oxygen octahedral size or the PDOS of the Ni. The localized electron on a Ni site resulting in a high-spin $Ni^{2+}$, where Hund's rules are favored over a Jahn-Teller distortion. This can be clearly seen for ¼ and ½ e⁻/Ni in Table S2. The octahedral distortions observed for ¾ and 1 e⁻/Ni are also constrained by the volume of the cell, which is not allowed to change in the ionic relaxation in these calculations.

**Table S1:**
Total energies, band gaps, lattice vectors and non-orthorhombic angle of the 20 atom/cell of the monoclinic structures. A, C, G and F are 10 atom/cell antiferromagnetic orderings of the monoclinic structure.

| Magnetic Order | Total Energy (eV) | Gap(eV) | a (Å) | b (Å) | c (Å) | β (°) |
|---|---|---|---|---|---|---|
| A | -134.80 | 0.54 | 5.300 | 5.724 | 7.481 | 90.007 |
| C | -134.64 | 0.43 | 5.270 | 5.812 | 7.441 | 90.011 |
| G | -134.59 | 0.75 | 5.278 | 5.818 | 7.421 | 90.005 |
| F | -135.01 | 0.30 | 5.328 | 5.553 | 7.600 | 90.031 |

**Table S2:** Magnetic moment of each of the four Ni and properties of the compassing oxygen octahedron in the doped SNO structures with G-type magnetic ordering. We elected to keep the overall volume of the calculations fixed as in reality the electron doping will not occur the same regularity present in the calculation. While the overall octahedral tilt pattern is not affected by electron doping, the tilt angles become more acute. When the lattice parameters are allowed to relax the overall volume increases; however, the octahedral volume increases that the tilt angles become even more acute.

| Property | $Ni_1$ | $Ni_2$ | $Ni_3$ | $Ni_4$ |
|---|---|---|---|---|
| **0 added electrons per Ni** | | | | |
| Magnetic Moment ($\mu_B$) | 0.92 | -0.92 | 0.92 | -0.92 |
| Octahedral Volume (Å³) | 10.62 | 10.62 | 10.62 | 10.62 |
| Ni-O Distances (Å) | 2.17, 1.91, 1.93 | 1.91, 2.17, 1.93 | 1.91, 2.17, 1.93 | 2.17, 1.91, 1.93 |
| Ni-O-Ni$_B$ Angle (°) (Ni$_B$) | 148.8 (Ni$_2$) | 148.6 (Ni$_3$) | 148.8 (Ni$_4$) | 148.6 (Ni$_1$) |
| **1/4 added electrons per Ni** | | | | |
| Magnetic Moment ($\mu_B$) | 0.81 | -0.84 | 0.86 | -1.65 |
| Octahedral Volume (Å³) | 10.46 | 10.48 | 10.39 | 11.76 |
| Ni-O Distances (Å) | 2.18, 1.92, 1.87 | 1.92, 2.17, 1.89 | 1.90, 2.05, 1.99 | 2.18, 2.03, 1.99 |
| Ni-O-Ni$_B$ Angle (°) (Ni$_B$) | 146.9, 146.8(Ni$_2$) | 145.6 (Ni$_3$) | 147.6, 148.0 (Ni$_4$) | 147.6 (Ni$_1$) |
| **1/2 added electrons per Ni** | | | | |
| Magnetic Moment ($\mu_B$) | 0.80 | -0.80 | 1.63 | -1.63 |

| Property | Ni$_1$ | Ni$_2$ | Ni$_3$ | Ni$_4$ |
|---|---|---|---|---|
| Octahedral Volume (Å$^3$) | 10.47 | 10.47 | 11.41 | 11.41 |
| Ni-O Distances (Å) | 2.18, 1.93, 1.87 | 1.93, 2.18, 1.87 | 2.04, 2.08, 2.03 | 2.08, 2.04, 2.03 |
| Ni-O-Ni$_B$ Angle (°) (Ni$_B$) | 145.5(Ni$_2$) | 144.6 (Ni$_3$) | 145.4(Ni$_4$) | 144.6 (Ni$_1$) |
| **3/4 added electrons per Ni** | | | | |
| Magnetic Moment (μ$_B$) | 0.75 | -1.61 | 1.62 | -1.62 |
| Octahedral Volume (Å$^3$) | 10.24 | 11.50 | 11.35 | 11.40 |
| Ni-O Distances (Å) | 2.09, 1.97, 1.88 | 2.05, 2.17, 1.96 | 2.06, 2.10, 1.98 | 2.06, 2.04, 2.04 |
| Ni-O-Ni$_B$ Angle (°) (Ni$_B$) | 143.6, 144.4(Ni$_2$) | 141.4 (Ni$_3$) | 144.5, 143.6 (Ni$_4$) | 142.8 (Ni$_1$) |
| **1 added electron per Ni** | | | | |
| Magnetic Moment (μ$_B$) | 1.61 | -1.61 | 1.61 | -1.61 |
| Octahedral Volume (Å$^3$) | 11.31 | 11.31 | 11.31 | 11.31 |
| Ni-O Distances (Å) | 2.09, 2.06, 1.98 | 2.06, 2.09, 1.98 | 2.06, 2.09, 1.98 | 2.09, 2.06, 1.98 |
| Ni-O-Ni$_B$ Angle (°) (Ni$_B$) | 142.5(Ni$_2$) | 139.3 (Ni$_3$) | 142.5 (Ni$_4$) | 139.3 (Ni$_1$) |
| **1 added electron per Ni (Lattice Parameters Relaxed, Fig. S4a, third panel)** | | | | |
| Magnetic Moment (μ$_B$) | 1.66 | -1.66 | 1.66 | -1.66 |
| Octahedral Volume (Å$^3$) | 13.78 | 13.78 | 13.78 | 13.78 |
| Ni-O Distances (Å) | 2.27, 2.11, 2.15 | 2.11, 2.27, 2.15 | 2.11, 2.27, 2.15 | 2.27, 2.11, 2.15 |
| Ni-O-Ni$_B$ Angle (°) (Ni$_B$) | 140.7(Ni$_2$) | 136.2 (Ni$_3$) | 140.7 (Ni$_4$) | 136.2 (Ni$_1$) |

## III. Adaptive Synaptic Plasticity Learning

In the Spiking Neural Network (SNN) simulations for digit recognition, we use the Leaky-Integrate-and-Fire (LIF) model[14,15] to simulate the membrane potential $V$ of a neuron as

$$\tau \frac{dV}{dt} = (E_{rest} - V) + g_e * (E_{exc} - V) + g_i * (E_{inh} - V)$$

where $E_{rest}$ is the resting membrane potential (-65 mV), $E_{exc}$ (0 mV) and $E_{inh}$ (-100mV) the equilibrium potentials of excitatory and inhibitory synapses, $\tau$ is the time constant (100 ms) and $g_e$ and $g_i$ are the conductances of excitatory and inhibitory synapses respectively. The LIF model causes $V$ to increase when pre-synaptic spikes are received and to otherwise decay exponentially. The post-neuron fires when V crosses the membrane threshold $V_{thresh}$ (-52 mV) and its membrane potential is reset to $V_{rst}$ (-65 mV). After each firing event, a refractory period (5 ms) ensues during which the post-neuron is inhibited from firing even if additional input spikes arrive.

Synapses are modeled by conductance changes[14,15] wherein the conductance increases by the synaptic weight, *w*, only upon the arrival of pre-neuronal spike. Otherwise, the conductance continues to decay exponentially. The dynamics of both inhibitory and excitatory conductance are simulated as

$$\tau_e \frac{dg_e}{dt} = -g_e, \quad \tau_i \frac{dg_i}{dt} = -g_i$$

where $\tau_e$ (1ms) or $\tau_i$ (2ms) are the time constants for the excitatory or inhibitory post-synaptic potential.

As discussed in Fig. S8, homeostasis is used to prevent a single neuron from dominating the spiking pattern. Specifically, each excitatory neuron's membrane threshold is not only determined by $V_{thresh}$ but by $V_{thresh}+\theta$, where $\theta$ is increased each time the neuron fires and then decays exponentially at an extremely slow rate. We use $\theta = 0.1$ and a very high decay time constant of $10^8$ ms in our simulations.

Each input image is presented for 350 ms. There is resting period of 150 ms before presenting a new input to allow all neuronal parameters to decay to the reset values (except for the adaptive membrane threshold, $V_{thresh}+\theta$). We note to the reader that we use identical parameters for neuron and synapse models, input encoding and input image presentation time as Diehl & Cook[15] for fair comparison of our ASP learning with standard STDP learning (Fig. 3b, Fig. S9). The standard STDP learning model is implemented using the power law weight dependent rule[15,16].

*ASP: Learning rules*

We examine the mathematical formulations for ASP to understand how the temporal dynamics dictate the plasticity that eventually enables the SNN to learn to forget as well as adapt to new patterns.

*Recovery Phase*

To improve simulation speed, the weight dynamics are computed using synaptic traces[17]. In ASP learning, the synapses keep track of three different kinds of traces corresponding to pre and post-synaptic neuron's spiking activity: a) Recent presynaptic trace ($Pre_{rec}$) that doesn't accumulate over time (only accounts for the most recent spike), b) Accumulative presynaptic ($Pre_{acc}$) trace that adds over time (accounts for the entire spike history of the presynaptic neuron for a given time period or epoch during which a particular pattern is presented to the SNN), c) Postsynaptic trace (*Post*) that accumulates over time based on the postsynaptic neuron's spiking activity. Each of the traces is evaluated as follows wherein the trace is increased when a spiking activity is observed, otherwise it decays exponentially:

$$Pre_{rec}(t) = \exp\left(-\frac{Pre_{rec}}{\tau_{rec}}\right); \; Pre_{rec} = 1 \; when \; t_{pre} occurs \quad (1)$$

$$Pre_{acc}(t) = \exp\left(-\frac{Pre_{acc}}{\tau_{acc}}\right); \; Pre_{acc} += 1 \; when \; t_{pre} occurs \quad (2)$$

$$Post(t) = \exp\left(-\frac{Post}{\tau_{post}}\right); \; Pre_{acc} += 1 \; when \; t_{post} occurs \quad (3)$$

Now, the time constant for decay of the accumulative pre-trace ($Pre_{acc}$) has to be larger than that of the recent pre-trace ($Pre_{rec}$) so that spike history can be appropriately added. In our simulations, $\tau_{acc} = 10\tau_{rec}$, $\tau_{post} = 2\tau_{acc}$. We adopt a modified version of the power-law weight dependent STDP model[14,15] to obtain the weight changes during the recovery phase (i.e. in presence of input stimulus) of ASP. When a postsynaptic spike arrives at the synapse, the weight change Δw is calculated based on the presynaptic trace ($Pre_{rec}, Pre_{acc}$)

$$\Delta w = \eta(t)(Pre_{rec} - offset) - \frac{k_{const}}{2^{Pre_{acc}}} \quad (4)$$

where η(t) is a time dependent learning rate that is inversely proportional to the post-synaptic trace value (*Post(t)* from Eqn. 3) at a given time instant. As the post-synaptic neuron in the excitatory layer starts spiking for a given input, the learning rate will decrease. This will ensure that a particular neuron retains and stably learns a particular input pattern. It also prevents the neuron from quickly adapting to a new pattern (or catastrophic forgetting). The offset ensures that the presynaptic neurons that rarely lead to firing of the postsynaptic neuron will become more and more disconnected (or the synaptic weight values will depress). In case of digit inputs, the black (or off) pixel region for a particular digit will become disconnected resulting in lowering of synaptic weight values corresponding to the pre-neurons in the lower pixel intensity region. In Eqn. 4, the first part represents the weight change (potentiation or depression) based on the most recent pre-synaptic spike (as with STDP). However, as seen earlier, erasure of memory traces is prominent with STDP as in its simplest form any pre/post spike pair will modify the synapse. Besides precise spike timings that identify the correlation between input patterns, learning rule should incorporate the significance of the inputs to modulate the weights. As the inputs are continually changed, an SNN (with fixed resources or size) should gradually forget obsolete data while retaining important information. Thus, input based significance driven learning would enable the SNN to learn in a stable-plastic manner in a dynamic environment.

The second part of Eqn. 4 quantifies the dependence of the weight change on the significance of the input pre-neuron. We define an input neuron to be significant if it has more frequent spikes. In that case, $Pre_{acc}$ value will be high that would eventually make the second term in Eqn. 4 less dominant for determining the final weight update. Thus, for more frequent input spikes at the pre-neuron, the weight update will be more prominent. Hence, the learning rule encompasses significance of the inputs with standard synaptic plasticity. It can be deduced that the prominent weights will essentially encode the features that are common to different input classes as the pre-neurons across those common feature regions in the input image will have higher firing activity. This eventually helps the SNN to learn more common features across different input patterns to obtain more generic representation of the data.

*Decay Phase*

The decay phase in ASP learning is activated in the absence of input stimulus i.e. when no spiking activity is observed at the synaptic terminals connecting pre and post neuron. It involves the forgetting of the weights for insignificant information to enable the SNN to learn new data without catastrophic forgetting or overlap of representations. As discussed earlier, the weights undergo an exponential decay towards a baseline value as

$$\tau_{leak} \frac{dw}{dt} = -\alpha w \quad (5)$$

where α is a decay constant and $\tau_{leak}$ is the time constant of decay. $\tau_{leak}$ is a time dependent quantity that is proportional to the post-synaptic trace value (Post(t) from Eqn. 3) and the membrane threshold value ( $v_{thresh} + \theta$ obtained from homeostasis behavior) at a given time instant. Now, it is desirable that the weights that have learnt a pattern should leak less in order to retain the learnt information. A neuron that has learnt a particular pattern will have a higher spiking activity (or higher post trace value Post(t)) that will increase the time constant of decay, $\tau_{leak}$. Higher $\tau_{leak}$ causes the weight to forget or leak less. Post(t) will be higher for an excitatory neuron that has learnt an input pattern that is recent and presented latest to the network. The overall leak rate can be defined as $\alpha/\tau_{leak}$ that decreases with increasing $\tau_{leak}$.

While Post(t) is indicative of how recent and latest the input pattern is, it does not account for the significance of the input pattern. We define significance in terms of number of times a particular pattern has been presented to an SNN. The membrane threshold (obtained from homeostasis) of a post-neuron is representative of the significance of the input pattern. A neuron that has learnt a given pattern will spike more when that pattern is presented several times to the network. An excitatory layer neuron's membrane threshold will be high only when it is firing more. Higher membrane threshold implies that the corresponding excitatory layer neuron has learnt a significant pattern. Hence, the SNN learns to forget insignificant older information while trying to retain more recent and significant, yet old, data using ASP.

A key aspect to note here is that the weight leak in the decay phase is based only on the post-neuron's spiking activity (and membrane threshold). All weights connected to a post-neuron in the excitatory layer will have the same decay time constant, $\tau_{leak}$ and hence show uniform leak dynamics during the decay phase. On the contrary, during recovery phase, the weight dynamics of each synapse will be different as it is determined by both the post and pre-neuronal spiking activity.

## IV. ASP-based 'Learning to forget' is compatible with other proposed neuromorphic device technologies

As discussed earlier, the dopant interaction with the perovskite lattice seen in experiment and studied by *ab-initio* dynamical simulations enables habituation-based plasticity. This is key to the perovskite's forgetting capability that motivates the ASP learning. In recent years, non-volatile and/or filamentary switch device elements including spin-based and memristors to emulate the behavior of neural systems have been proposed[18-26]. While our correlated perovskite can emulate forgetting similar to the animal world, other non-volatile devices can be made to forget or leak their conductance by applying electrical pulses. Our proposed ASP can therefore be synergistic with those devices as well. Thus, ASP can be incorporated with a broad range of programmable devices to construct robust self-adaptive artificial neural systems for dynamic environments.

## V. AIMD Movie

*Ab initio* molecular dynamics simulations showing migration of proton in H-doped monoclinic SNO crystal at 300 K. The proton hops from one O atom to another neighboring O atom within the $NiO_6$ octahedron in a facile manner (see Fig. 2e for details on the activation barriers). The Ni, O, Sm and H are depicted as green, red, yellow, and blue spheres respectively. For the sake of clarity, only the hydrogen, and the Ni/O atoms belonging to the two $NiO_6$ octahedra closest to the hydrogen are shown as large spheres; the atoms far away from the hopping phenomena are depicted with small translucent spheres. Our AIMD simulations at 300 K at various H doping levels show that the SNO lattice monotonically expands with addition of hydrogen approaching lattice expansion of ~5% for 1 H per unit cell of SNO.